\documentclass[twocolumn,showpacs,amsmath,amssymb,prl,superscriptaddress]{revtex4}
\usepackage{graphicx}
\usepackage{bm}
\usepackage{dcolumn}

\begin{document}
\title{Microscopic origin of large negative magneto-electric coupling in Sr$_{1/2}$Ba$_{1/2}$MnO$_{3}$}

\author{Gianluca Giovannetti}
\affiliation{CNR-IOM-Democritos National Simulation Centre and International School
for Advanced Studies (SISSA), Via Bonomea 265, I-34136, Trieste, Italy}
\affiliation{Institute for Theoretical Solid State Physics, IFW-Dresden, PF 270116, 01171 Dresden, Germany}
\author{Sanjeev Kumar}
\affiliation{Indian Institute of Science Education and Research(IISER) Mohali, Knowledge City, Sector 81, 
Mohali 140 306, India}
\author{Carmine Ortix}
\affiliation{Institute for Theoretical Solid State Physics, IFW-Dresden, PF 270116, 01171 Dresden, Germany}
\author{Massimo Capone}
\affiliation{CNR-IOM-Democritos National Simulation Centre and International School
for Advanced Studies (SISSA), Via Bonomea 265, I-34136, Trieste, Italy}
\author{Jeroen van den Brink}
\affiliation{Institute for Theoretical Solid State Physics, IFW-Dresden, PF 270116, 01171 Dresden, Germany}

\date{\today}

\begin{abstract}
With a combined {\it {\it ab initio}} density functional and model Hamiltonian approach we establish that in the recently discovered multiferroic phase of the manganite Sr$_{1/2}$Ba$_{1/2}$MnO$_{3}$ the polar distortion of Mn and O ions is stabilized via enhanced in-plane Mn-O hybridizations.
The magnetic superexchange interaction is very sensitive to the polar bond-bending distortion, and we find that this dependence directly causes a strong magnetoelectric coupling. This novel mechanism for multiferroicity is consistent with the experimentally observed reduced ferroelectric polarization upon the onset of magnetic ordering.
\end{abstract}

\maketitle

Multiferroic materials are ideal  candidates for the realization and practical use of strong magnetoelectric effects~\cite{Hill,Cheong}. The scarcity of  actual materials that are magnetic ferroelectrics appears to be related to the competition between the conventional mechanism of ferroelectric cation off-centering, which requires empty $d$-orbitals, and the formation of magnetic moments which requires partially filled $d$-orbitals~\cite{Hill,Cheong}. A concomitance of magnetism and ferroelectricity then has to rely on more subtle microscopic coupling mechanisms, driven by spin-orbit coupling in the form of Dzyaloshinskii-Moriya interactions~\cite{Katsura} or exchange-striction~\cite{Arima}. The recently synthesized manganite Sr$_{1/2}$Ba$_{1/2}$MnO$_{3}$ however defeats the generic incompatibility of a cation both having a magnetic moment and being ferroelectrically displaced. This system is a classic example of a material in which charge, spin, lattice and orbital degrees of freedom are strongly coupled, giving in this particular case rise to a strong magnetoelectric (ME) effect, the origin of which we set out to clarify here. 

For doing so, the methods from modern {\it ab initio} bandstructure theory are powerfull tools -- very helpful not only in predicting new multiferroic 
materials, but also in understanding the underlying mechanisms for magnetoelectric couplings. The computed values of macroscopic polarization $P$ agree exceptionally well with those observed experimentally~\cite{Picozzi1,125,CuO,Stroppa,Yamauchi,Baettig}. In the last few years several {\it ab initio} calculations have pointed out the possible ferroelectric  state with large polarization for AMnO$_3$, where A is an alkaline earth element. 
The proposed mechanism is  based on off-centering of Mn$^{4+}$ ions stabilized via a charge-lattice coupling of Peierls type~\cite{Bhattacharje,Rondinelli,Lee}.
The problems in synthesizing such a material with predicted ferroelectricity has very recently been overcome:  last year Sr$_{1/2}$Ba$_{1/2}$MnO$_{3}$ (SBMO) has  been reported to support a ferroelectric phase via the  off-centering of magnetic Mn$^{4+}$ ion in conjunction with a perovskite tetragonal structure~\cite{Sakai}. The onset of the low-temperature long-range antiferromagnetic (AFM) ordering strongly reduces the polarization indicating a large magnetoelectric effect~\cite{Sakai}.  The AFM order, in other words, does not support ferroelectricity, but it neither completely destroys it. This special feature of SBMO opens a new avenue for the quest of materials with strong ME effects, where the search need not be restricted to systems in which FE and magnetism mutually stabilize each other.

Here we establish with a combination of first-principles calculations and a model Hamiltonian analysis that the ferroelectric polarization mainly arises from a polar distortion of Mn and O ions caused by an enhanced in-plane Mn-O hybridization. Since the magnetic superexchange interaction strongly depends on this distortion, a strong and novel type of magnetoelectric coupling arises. This ME coupling is negative in the sense that the ferroelectric polarization is not promoted by magnetism, but rather reduced by it, which renders antiferromagnetic ordering and ferroelectricity strongly coupled.

We first present the results of our first-principles calculations based on density functional theory (DFT)\cite{DFT} using the generalized gradient approximation (PBE)\cite{PBE} and including correlation effects within the DFT+U scheme \cite{DFTplusU} as implemented in VASP\cite{VASP}.
We use on-site Coulomb and exchange parameters U$=$3.0 and 4.5 eV and J$_H=$1.0 eV on the manganese d-orbitals. In the projector augmented wave scheme \cite{PAW} the cut-off for the plane-wave basis set was chosen as 400 eV and a 8$\times$8$\times$8 mesh was used for the Brillouin-zone sampling. To calculate the electronic contribution to the spontaneous polarization we use the Berry-phase method developed by King-Smith and Vanderbilt \cite{Vanderbilt}. In the calculations the in-plane lattice constant is taken as $a=$3.85 \AA. For the interplane distance we consider the two values $c/a=$1.005, 1.01, which are experimentally determined for the SBMO at different temperatures~\cite{Sakai}.

\begin{figure}
\includegraphics[width=.85\columnwidth,angle=0]{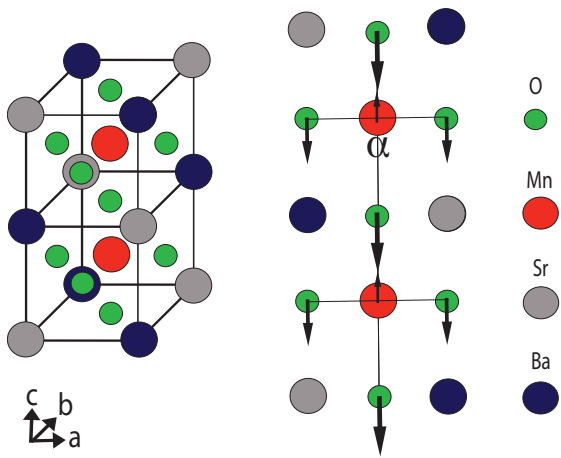}
\caption{(Color online) Schematic view of SBMO unit cell and displacements of Mn and O sites in AFM magnetic structure. Arrows indicate the relative atomic ferroelectric 
displacements.}
\label{fig1}
\end{figure}

For all the above parameters SBMO is safely in an AFM (G-type) insulating state with a band gap of $\sim$ 0.4 eV with Mn magnetic moments  $M \sim$ 2.6 $\mu_b$, in agreement with previous DFT calculations on CaMnO$_3$ and SrMnO$_3$ \cite{Rondinelli,Lee,Picozzi2} and experiments on SrMnO$_3$ \cite{Takeda}. The valence band is predominantly majority-spin Mn t$_{2g}$ and O 2p characther with strong $p-d$  hybridization while the conduction band is formed by Mn e$_g$ orbital and empty minority t$_{2g}$ states, which is consistent with Mn$^{4+}$ in octahedral crystal field.
To find the energetically most stable configuration we relax the ions performing structural optimization in a 40 atoms 2$\times$2$\times$2 unit cell. We start the relaxation from a checkerboard arrangement of the Sr, Ba ions and check that our results do not depend on this assumption.

\begin{figure}
\includegraphics[width=.99\columnwidth,angle=0]{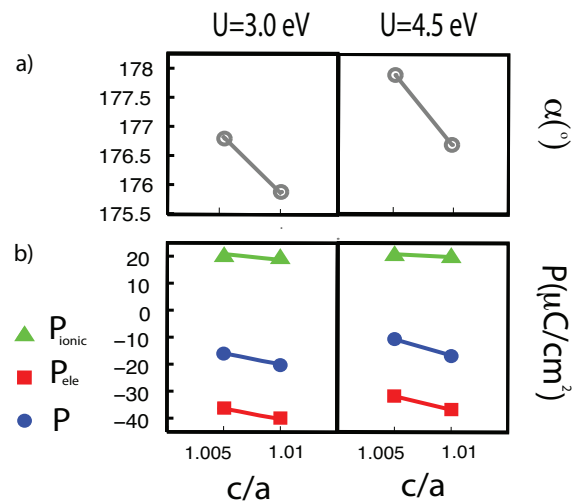}
\caption{(Color online) a),b) Values of $\alpha$ and spontaneous polarization $P$, $P_{ele}$, $P_{ionic}$ 
as function of the ratio $c/a$ for $U$=3.0 ev and 4.5 eV respectively.}
\label{fig2}
\end{figure}

Even if the initial ionic structure belongs to space group I4/mmm (No. 139) which is centrosymmetric, for all our parameters the relaxed structure belongs to the space group I4mm (No. 107) and it breaks inversion symmetry.  The polar ionic displacements associated with the reduced symmetry are shown schematically in Fig. \ref{fig1}. The O-Mn-O angle $\alpha$ (see Fig. \ref{fig1}) which is 180$^o$ for ideal centrosymmetric structure with $c/a=1$, is reduced in agreement with the experimentally determined low symmetry structure \cite{Sakai}. The deviation of $\alpha$ from 180$^o$ as function of $c/a$ is shown in Fig. \ref{fig2}.
We now analyze the effect of this reduced angle on the ferroelectric polarization. We first notice that the electronic contribution to the spontaneous polarization $P$ evaluated in the centrosymmetric structures (I4/mmm) for the AFM ground state is zero, meaning that the polar state is not magnetically driven. Indeed at different values of the ratio $c/a$ it is the covalent bond formation upon ionic displacements between e$_g$ orbitals of Mn and $p$ orbitals of apical O ions to determine the  stabilization of the ferroelectric state \cite{Rondinelli}.

On the other hand in the relaxed state, the large polar displacements of the apical O ions along the $c$ lattice direction result in the formation of dipolar pairs between manganese and oxygen (see Fig. \ref{fig1}) and to a state  similar to a bond-centered charge density wave \cite{LCMO}. At low temperature, when the system orders antiferromagnetically, this enables the practical realization of a peculiar and atypical multiferroic state. This is shown by the results for the polarization $P$, whose electronic and ionic contributions ($P_{ele}$, $P_{ionic}$)  are plotted in Fig. \ref{fig2}. 
Increasing the ratio $c/a$ the magnitude of the electronic contribution $P_{ele}$ increases and that of the ionic contribution $P_{ionic}$ decreases. Since the two are opposite in sign the total polarization $P=P_{ele}+P_{ionic}$  increases as a function of $c/a$. The calculated value of $P$ agrees with the experimental value of 13.5 $\mu$C/cm$^2$ for single domain \cite{Sakai}. 
This physical result does not depend strongly on the structural and interaction ($U$, $J_H$) parameters, but the quantitative description of the  ferroelectric instability in SBMO should of course depend on the actual values of these parameters, as has been found to be the case in other Mn based multiferroic materials \cite{125}. In particular at larger U the magnetic moment increases and  the ferroelectric tendency decreases as the 
angle $\alpha$ gets closer to 180$^{\circ}$(see Fig. \ref{fig2}). 

\begin{figure}
\includegraphics[width=.99\columnwidth,angle=0]{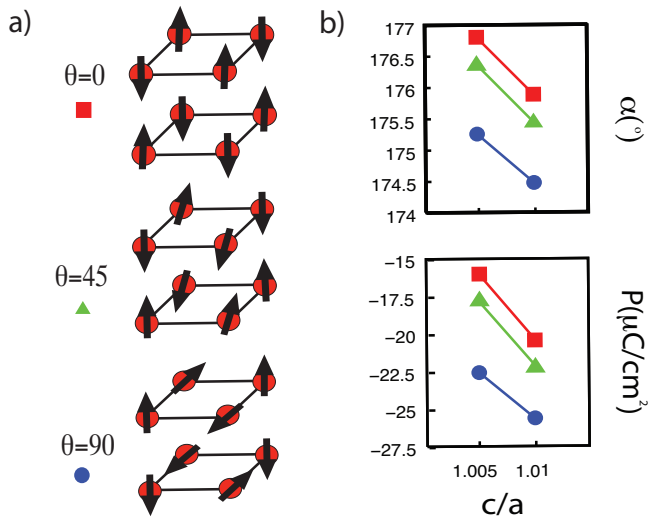}
\caption{(Color online) a) Schematic view of non-collinear magnetic structures having Mn spins with 
angle $\theta = 0^o, 45^o ,90^o$; b) Values of $\alpha$ and 
spontaneous polarization $p$ as function of the ratio $c/a$ for different values of $\theta$ at U$=$3.0 eV and J$_H=$1.0 eV.}
\label{fig3}
\end{figure}

The calculated  polarizations show that the FE order is not driven by the magnetic order, but yet the two are strongly coupled. 
This counterintuitive situation arises, as we will show next, from
the ferroelectric transition in SBMO being driven by Mn and O displacement and the mechanism of the suppression of tetragonal distortion below T$_{N}$ \cite{Sakai} being due to the subsequent strong change in the superexchange interactions betweens Mn spins \cite{GoodenoughKanamori}.
The 180$^o$ O-Mn-O bonds are energetically favored by the antiferromagnetic coupling \cite{GoodenoughKanamori} then in the ferroelectric state the off-centering of Mn ions, which is in favor of the inset of double exchange interactions, gets suppressed with a net decreasing of the ferroelectric polarization \cite{BrinkKhomskii}.
The effect of the magnetism on the ferroelectric distortions can be captured by performing calculations with non-collinear magnetic structures having Mn spins with angle $\theta$ ranging from 0$^o$ (G-type) to 90$^o$ (see Fig. \ref{fig3}a) to control how the superexchange interactions along the O-Mn-O bonds changes $P$. At each angle $\theta$ the lattice structure is relaxed and the sum of electronic and ionic contributions to the ferrolectric polarization is evaluated (see Fig. \ref{fig3}b).
%
Increasing the angle $\theta$ between the spins reduces the superexchange interations. The Mn-O-Mn angle $\alpha$ decreases with a resulting larger Mn off-centering which stabilizes the ferroelectric polarization. We observe that the magnetic order alters both electronic and ionic contributions to the polarization via a change in $\alpha$: the magnetism is thus coupled to the lattice and the latter is coupled to the polarization.
Increasing the superexchange interactions causes the magnetic structure to drive the lattice towards a recovery of the a centrosymmetric arrangement.

To flesh out the microscopic origin of the ferroelectric instability we set up a model Hamiltonian, based on the bandstructure results and taking into account the different magnetic exchange interactions and the coupling of the electrons to the lattice:
\begin{eqnarray*}
H =&&
- \sum_{i, \gamma, \sigma}
t^{\gamma}(u_i) \left ( d^{\dagger}_{i, \sigma} p^{\gamma}_{i+\gamma, \sigma} + H.c. \right ) 
+ \sum_i \Delta_{pd} ~ d^{\dagger}_{i, \sigma} d^{}_{i, \sigma}  
\cr
&&
 - J_H \sum_{i} {\bf S}_i \cdot {\mbox {\boldmath $\sigma$}}_{i}
+ J_s \sum_{i, \gamma} {\bf S}_i \cdot {\bf S}_{i+\gamma} + K_s \sum_{i} {u_i}^2. 
\end{eqnarray*}
Here, $d^{}_{i, \sigma}$ ($p^{\gamma}_{i, \sigma}$) and $d^{ \dagger}_{i, \sigma}$  ($p^{\dagger}_{i, \sigma}$) are the annihilation and creation operators for Mn-$d$ (O-$p^{\gamma}$) electrons with spin $\sigma = \uparrow, \downarrow$.  ${\bf S}_i$ are the localized $t_{2g}$ spins ($S=3/2$),  which in this study are treated classically and coupled antiferromagnetically via $J_s$. $u_i$ are the off-centering distortions of Mn ions along the c-axis and $K_s$ denotes the stiffness energy associated with these distortions.
$t^{\gamma}(u_i)$ denote the distortion-dependent hopping amplitudes between
$d_{3z^2-r^2}$ and $p^{\gamma}_z$ orbitals along $\gamma$ direction ($\gamma = x,y,z$). Note that $\gamma$ denotes the direction in real space and not the character of the $p$ orbitals. The ${\mbox {\boldmath $\sigma$}}_{i}$ denote the electronic spin operator defined as, ${\sigma}^{\mu}_{i}= \sum_{\sigma \sigma'} d^{\dagger}_{i\sigma} \tau^{\mu}_{\sigma \sigma'} d_{i \sigma'}$,
where $\tau^{\mu}$ are the Pauli matrices. $\Delta_{pd}$ is the on-site energy  difference between Mn-$d_{3z^2-r^2}$ and O-$p_z$ levels. 

In the model Hamiltonian the ionic displacements are restricted to the $c$-axis direction, as observed in the experiments and verified in our bandstructure calculations. In principle the O ions are easier to displace, however a combination of O displacements and Mn displacements can be modeled as a net off-centering displacement of the Mn ions along with an overall change in the lattice $c$ parameter. Here, we model the effective displacements via the off-centering $u_i a$ of the Mn ions, where $a$ is the Mn-Mn lattice spacing. We consider only $d_{3z^2-r^2}$ orbital as the one that can hybridize with the O-$p_z$ levels, since the planar orbitals $d_{x^2-y^2}$ have zero overlap with the O-$p_z$. 

If Mn ions are located at the center of O$_6$ octahedra then the hopping between $d_{3z^2-r^2}$ and O-$p_z$ is non-zero only along z-axis and is given by $t_0 = (pd\sigma$). However, if Mn ions are off-centered by a small displacement  they lead to a finite in-plane hopping which can be calculated from the Slater-Koster tables as 
$t^{x/y} = n(n^2 - (l^2+m^2)/2) (pd\sigma) + \sqrt{3} n (l^2+m^2) (pd\pi), $
where $l,m,n$ are the direction cosines from O to Mn \cite{SlaterKoster}. Taking only the $pd\sigma$ contribution one can write the hopping integral in terms of the Mn-O-Mn angle $\alpha$ as, 
$ t^{x/y} = \sin (\alpha/2)(\sin^2(\alpha/2) - \cos^2(\alpha/2)/2) (pd\sigma)$. Rewriting the trigonometric functions in terms of the the distortions, we get
to leading order in the 
distortion $u_0$,  $t^{x/y} \sim -2u_0 ~(pd\sigma)$. The next order term is $O(u_0^3)$ which can be safely ignored. Naturally $t_{pd}^z$ is also modified via a Peierls type term with the hopping between longer and shorter bonds given by $t^z_{\pm} = (1 \pm gu)t_0$. 

Just as the off-centering of Mn ions affects the hopping parameters $t_{pd}^{\gamma}$, it also affects the value of  $J_s$ via the Mn-O-Mn bond angle. $J_s$ is maximum at $\alpha = \pi$ and is reduced by any deviation.  The leading order change in the Taylor expansion around the point $\alpha = \pi$ is $O(\delta \alpha^2)$.  Therefore, for small deviations we can model the distortion-dependence as $J_s^{x/y} = -J_0 \cos (\alpha)$. In principle $J_s^z$ is also affected since the distances Mn-O$_1$ and Mn-O$_2$ for the two apical oxygens become unequal, but this dependence does not affect the physical picture.

\begin{figure}
\includegraphics[width=.95\columnwidth,angle=0, clip=true]{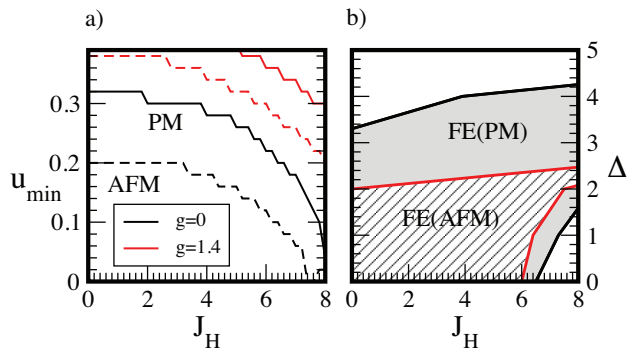}
\caption{(Color online)  a) $u_{min}$ as a function of $J_H$ for paramagnetic and 
antiferromagnetic spin configurations for $g=0$ and $g=1.4$ . AFM state 
leads to a reduction in FE distortion. b) Phase diagram in $\Delta$-$J_H$ phase space showing the 
regions of FE stability.}
\label{fig4}
\end{figure}

Given a specific configuration of lattice distortions and $t_{2g}$ spins, one can easily diagonalize the electronic problem numerically on finite lattices. 
We use K$_s$ =15 and J$_s$ = 0.1 while calculating the dependencies on the other parameters. Our focus is to explore the possibility of a FE state in in both the nonmagnetic and AFM phase. Therefore, rather than performing a lengthy minimization of the energy as a function of classical spin and lattice variables we compare the total energy of only the relevant configurations. We use the magnitude of off-centering distortions $u(i) \equiv  u_0$ as a variational parameter and determine the distortions $u_{min}$ that correspond to the lowest total energy. A non-zero value of $u_{min}$ is the hallmark of a  FE state. 
The onset of antiferromagnetism leads to a reduction in the tendency to form a FE state, which is reflected in a reduced value of $u_{min}$ for AFM order shown in Fig. \ref{fig4} a).

The results of model calculations are summarized in the $\Delta - J_H$ phase diagram in Fig. \ref{fig4}b. 
The phase diagram shows that the FE phase is stabilized over a wide range of parameter space when the system is in PM state. The presence of AFM order shrinks the regions of stability of the FE order, and in general the AFM order reduces the value of FE  polarization for all parameter values. We present the phase diagram for $g=0$, which shows that the mechanism for FE ordering does not depend on the Peierls type electron-lattice coupling. However, the presence of a non-zero $g$ further stabilizes the FE order.

In conclusion, by combining different theoretical approaches we highlight the intricate interrelationship between 
magnetic and ferroelectric orderings in recently discovered multiferroic phase of 
Sr$_{1/2}$Ba$_{1/2}$MnO$_{3}$ \cite{Sakai}. The new mechanism at play relies on the distortion dependent in-plane 
hopping between Mn and O sites and strongly depends on the onset of the magnetic order. 
Via an interplay between charge, spin, lattice and orbital degrees of freedom this leads to the experimentally observed magnetically suppressed ferroelectricity.  
This type of strong magneto-electric coupling being present in Sr$_{1/2}$Ba$_{1/2}$MnO$_{3}$ opens new routes for the search of multiferroic materials different from other Mn based oxides such as RMnO$_3$ and RMn$_2$O$_5$ \cite{Cheong}, going beyond the requirement of the magnetic ordering breaking the inversion symmetry, whereby it causes a ferroelectric instability. 

This work is supported by CINECA who allocated computer time. 
M.C. and G.G acknowledge financial support by the European Research Council under FP7/ERC Starting Independent Research Grant ``SUPERBAD" (Grant Agreement n. 240524). We thank R.O. Kuzian, S.-L. Drechsler and J. Malek for stimulating discussions.


\begin{thebibliography}{99}
\bibitem{Hill} N.A. Hill, J. Phys. Chem. B 104, 6694 (2000).
\bibitem{Cheong} S.W. Cheong {\it et. al.} , Nature Mat. 6, 13 (2007).
\bibitem{Katsura} H. Katsura {\it et. al.}, Phys. Rev. Lett. 95, 057205 (2005).
\bibitem{Arima} T. Arima {\it et. al.}, Phys. Rev. Lett. 96, 097202 (2006).
\bibitem{Picozzi1} S. Picozzi {\it et. al.}, Phys. Rev. Lett. 99, 227201 (2007).
\bibitem{125} G. Giovannetti {\it et. al.}, Phys. Rev. Lett. 100, 227603 (2008).
\bibitem{CuO} G. Giovannetti {\it et. al.}, Phys. Rev. Lett. 106, 026401 (2011).
\bibitem{Stroppa} A. Stroppa {\it et. al.}, New J. Phys. 12 093026 (2010).
\bibitem{Yamauchi} K. Yamauchi {\it et. al.}, Phys. Rev. B 84, 165137 (2011).
\bibitem{Baettig} P. Baettig {\it et. al.}, Phys. Rev. B 72, 214105 (2005).
\bibitem{Bhattacharje} S. Bhattacharje {\it et. al.}, Phys. Rev. Lett. {\bf 102}, 117602 (2009).
\bibitem{Rondinelli} J. M. Rondinelli {\it et. al.},  Phys. Rev. B {\bf 79}, 205119 (2009).
\bibitem{Lee} J. H. Lee {\it et. al.}, Phys. Rev. Lett. {\bf 104}, 207204 (2010).
\bibitem{Sakai} H. Sakai {\it et. al.}, Phys. Rev. Lett. {\bf 107}, 137601 (2011).
\bibitem{DFT} P. Hohenberg and W. Hohn, Phys. Rev. {\bf 136} B864 (1964); W. Kohn and L. J. Sham, Phys. Rev. {\bf 140} A1133 (1965).
\bibitem{PBE} J.P. Perdew {\it et. al.}, Phys. Rev. Lett. {\bf 77}, 3865 (1996).
\bibitem{DFTplusU} A. I. Liechtenstein {\it et. al.}, Phys. Rev. B {\bf 52}, R5467 (1995).
\bibitem{VASP} G. Kresse and J. Furthmuller, Phys. Rev. B {\bf 54}, 11 169 (1996); G. Kresse and J. Furthmuller, Comput. Mater. Sci. {\bf 6}, 15 (1996).
\bibitem{PAW} G. Kresse and D. Joubert, Phys. Rev. B {\bf 59}, 1758 (1999).
\bibitem{Vanderbilt} R.D. King-Smith {\it et. al.}, Phys. Rev. B {\bf 47}, 1651 (1993); D. Vanderbilt {\it et. al.}, Phys. Rev. B {\bf 48}, 4442 (1994).
\bibitem{Picozzi2} S. Picozzi {\it et. al.},  Phys. Rev. B {\bf 75}, 094418 (2007).
\bibitem{Takeda} T. Takeda {\it et. al.}, J. Phys. Soc. Jpn. {\bf 37}, 275 (1974).
\bibitem{LCMO} G. Giovannetti {\it et. al.}, Phys. Rev. Lett. 103, 037601 (2009).
\bibitem{GoodenoughKanamori} J. B. Goodenough, Phys. Rev. 100, 564 (1955); J. Kanamori, J. Phys. Chem. Solids 10, 87 (1959).
\bibitem{BrinkKhomskii} J. van den Brink {\it et. al.}, J. Phys.: Condens. Matter 20, 434217 (2008).
\bibitem{SlaterKoster} J. C. Slater and G. F. Koster, Phys. Rev. 94, 1498(1954).
\end{thebibliography}
\end{document}